# A post mortem analysis of the strain-induced crystallization effects on fatigue of elastomers

B. Ruellan[1,2,3], J.-B. Le Cam[1,2], E. Robin[1,2], I. Jeanneau[2,3], F. Canévet[2,3], G. Mauvoisin[4] and D. Loison[1]

1. Univ Rennes, CNRS, IPR (Institut de Physique de Rennes) - UMR 6251, F-35000 Rennes, France
2. LC-DRIME, Joint Research Laboratory, Cooper Standard - Institut de Physique UMR 6251, Campus de Beaulieu, Bât. 10B, 35042 Rennes Cedex, France.
3. Cooper Standard France, 194 route de Lorient, 35043 Rennes - France.
4. Laboratoire de Génie Civil et Génie Mécanique EA 3913, IUT-Université de Rennes 1, France.

## ABSTRACT

Natural rubber (NR) is the most commonly used elastomer in the automotive industry thanks to its outstanding fatigue resistance. Strain-induced crystallization (SIC) is found to play a role of paramount importance in the great crack growth resistance of NR [1]. Typically, NR exhibits a lifetime reinforcement for non-relaxing loadings [2-3]. At the microscopic scale, fatigue striations were observed on the fracture surface of Diabolo samples tested in fatigue. They are the signature of SIC [2,4,5]. In order to provide additional information on the role of SIC in the fatigue crack growth resistance of NR, striations are investigated through post-mortem analysis after fatigue experiments using loading ranging from -0.25 to 0.25. No striation was observed in the case of tests performed at 90°C. This confirms that the formation of striation requires a certain crystallinity level in the material. At 23°C, two striation regimes were identified: small striation patches with different orientations (Regime 1) and zones with large and well-formed striations (Regime 2). Since fatigue striations are observed for all the loading ratios applied, they are therefore not the signature of the reinforcement. Nevertheless, increasing the minimum value of the strain amplified the striation phenomenon and the occurrence of Regime 2.

Keywords: natural rubber, strain-induced crystallization, fatigue, striation

## INTRODUCTION

Elastomers are widely used in many applications for their extraordinary mechanical properties. Typically, they present damping abilities and a high fatigue resistance. The fatigue resistance of elastomers was investigated as soon as 1940 by the pioneer work by Cadwell *et al.* [2], Fielding [6] and 20 years later by Beatty [7]. In the case of natural rubber (NR), the strong impact of the loading ratios on the fatigue properties was demonstrated. For non-relaxing loadings (*i.e.* R > 0), it corresponds to a decrease in fatigue crack growth rate and an increase in tearing energy below which no crack growth occurs [1]. Since this result has not been observed for non-crystallizable rubbers, it is commonly attributed to strain-induced crystallization (SIC). Nevertheless, the mechanisms responsible for this reinforcement are not fully understood and remain not clearly established, which is an obstacle for lifetime prediction. The presence of certain morphological patterns, namely wrenchings and striations [4], were assumed to be due to SIC. Therefore, their presence on the failure surface of NR samples should be the signature of the SIC effects in the crack growth resistance at the microscopic scale. On the one hand, the mechanisms of wrenching formation were established by Le Cam *et al.* [8], it revealed the role of SIC in the process of formation: highly crystallized ligaments form between elliptical zones where the crack propagates, which delay the crack propagation. On the other hand, only few studies investigated striations [4,5,9,10]. Therefore, the role of SIC in their formation is not clearly understood, especially for non-relaxing loadings, where the lifetime reinforcement occurs.

Fatigue striations were investigated in four studies [4,5,9,10]. The state of the art on the experiments carried out is given in [11]. A summary of this state of the art is given hereafter. First of all, the experimental conditions used in the literature to investigate fatigue striation strongly differed from one study to another. Therefore, comparing them or drawing general conclusions on fatigue striations seems complicated. The results obtained can be presented according to the sample geometry, since it was found to strongly influence the fatigue striation phenomenon.

*(i)* Plane samples

The study by Munoz et al. [10] is the one presenting the largest diversity of fatigue striations. In this study, the fatigue striation morphologies were classified according to the crack growth rate. Three striation morphologies T1, T2 and T3 were identified, they refer to morphological groups III, IV and IV in [10], respectively. They stand for striations patches whose orientation differs from one patch to another, well defined striations regrouped in different zones whose orientation also differs slightly and well-defined striations that people the sample thickness, respectively. The author showed that the occurrence of fatigue striation from typology T1 to T3 goes along with an increase in crack growth rate.

*(ii)* Diabolo samples

The crack first initiates around a defect that concentrates stress. Then, it propagates by forming small wrenchings describing an ellipsoidal zone around a defect [5]. The wrenchings exhibit a similar morphology as those observed in the case of plane samples. As the crack propagates, striations eventually form before the sample breaks totally. It is to note that striations have also been observed in the case of non-relaxing loadings in [5], but damage at the macroscopic scale was quite different from that described here, since the failure occurred in a region under the insert. At the microscopic scale and on a morphological point of view, striations are separated by a distance ranging from 10 to 100 µm in [9] and from 15 to 160 µm as the crack growth rate increases in [10].

As a conclusion, only a few studies investigated fatigue striations while they could bring new information on the role of SIC on the lifetime reinforcement under non-relaxing loadings. Furthermore, the studies differed in terms of material, loading condition and sample geometry, which makes difficult any generalization to NR. Additional fatigue tests have therefore to be carried out at different loading ratios and temperatures. The next section presents these new fatigue tests. The experimental setup is first provided. Then, results are given and discussed. Concluding remarks close the paper.

## EXPERIMENTAL SETUP

*Material and sample geometry*

The material considered here is a carbon black filled natural rubber (*cis*-1,4 polyisoprene) vulcanised with sulphur. Samples tested are Diabolo samples.

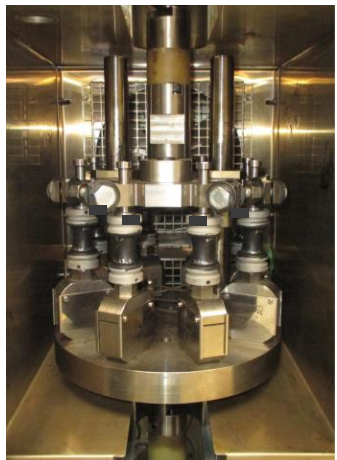

Figure 1: experimental setup.

*Loading conditions*

The fatigue tests are performed with a uni-axial MTS Landmark equipped with a homemade experimental apparatus presented in Figure 1. This apparatus enables us to test simultaneously and independently eight Diabolo samples, which compensates the fatigue tests duration and the dispersion in the fatigue lifetimes.

The tests are performed under prescribed displacement. The corresponding local deformation is calculated by finite element analysis (FEA) at the sample's surface in the median zone. In order to investigate the effect of temperature on the striation and fatigue properties, a Servathin heating chamber is used. The frequency is chosen in such a way that the global strain rate $\dot{\varepsilon}$ is kept constant, ranging between 1.8 and 2.4 to 2 $s^{-1}$ for a given loading condition. In practice, the frequency ranges between 1 and 4 Hz, which limits self-heating so that no thermal damage is added to the mechanical one. Four different loading ratios $R_\varepsilon = \frac{\varepsilon_{min}}{\varepsilon_{max}}$ are used: -0.25; 0; 0.125 and 0.25. One can recall that loading ratios inferior, equal and superior to zero correspond to tension- compression, repeated tension and tension-tension, respectively.

*Scanning Electron Microscopy*

Second electrons images of diabolo fracture surfaces are recorded with a JSM JEOL 7100 F scanning electron microscope (SEM). In addition, the SEM is coupled with an Oxford Instrument X Max Energy Dispersive Spectrometer of X-rays (EDS) and an Aztzec software in order to determine the surface fracture composition, especially in the crack initiation zone. The fracture surfaces to be analysed are previously metallized by vapour deposition of an Au-Pd layer.

## RESULTS AND DISCUSSIONS

The failure surface was generally similar to the one described by Le Cam and co-workers [4,5], *i.e.* the formation of fatigue striation follows the formation of wrenchings and precedes the final ligament corresponding to a catastrophic failure. Fatigue striations are first described according to their typologies. Then, the effects of the loading on the occurrence and typology of the fatigue striation are discussed.

*Typology of fatigue striations*

The analysis of the failed surfaces suggests fatigue striations form according to two regimes (see Figure 2):

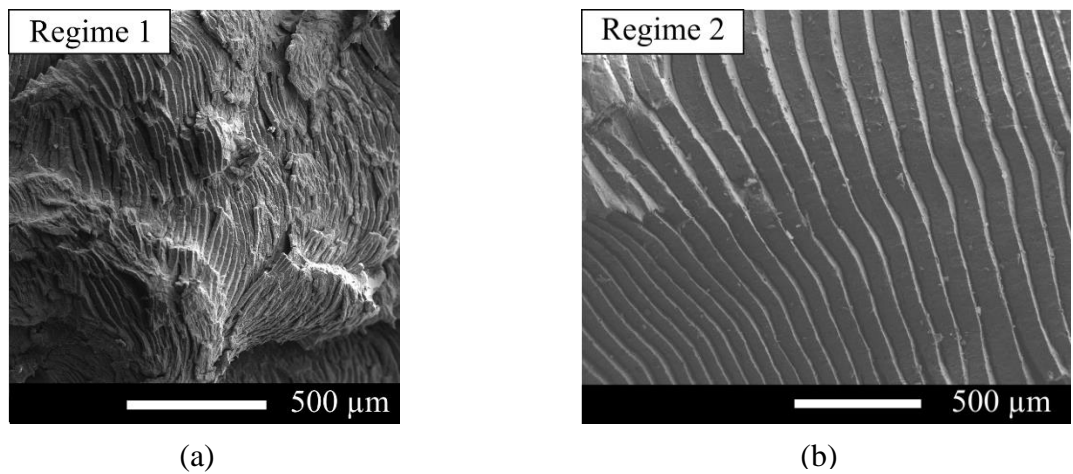

Figure 2: SEM images of the two striation regimes (a) Regime 1 and (b) Regime 2.

- Regime 1 corresponds to striation patches (see Figure 2(a)) whose orientation can slightly deviate from the crack propagation direction. Regime 1 corresponds to the transition between full wrenching zone and full striation zones. The loading can be considered as moderate.
- Regime 2 stands for well-formed striations (see Figure 2 (b)) and wrenchings no longer form. It corresponds to the striation morphology identified in the literature in Diabolo samples [4,5,9]. Since Regime 2 is observed closed to the final ligament, the local stress can be considered as severe. It is to note that Regime 1 always precedes Regime 2.

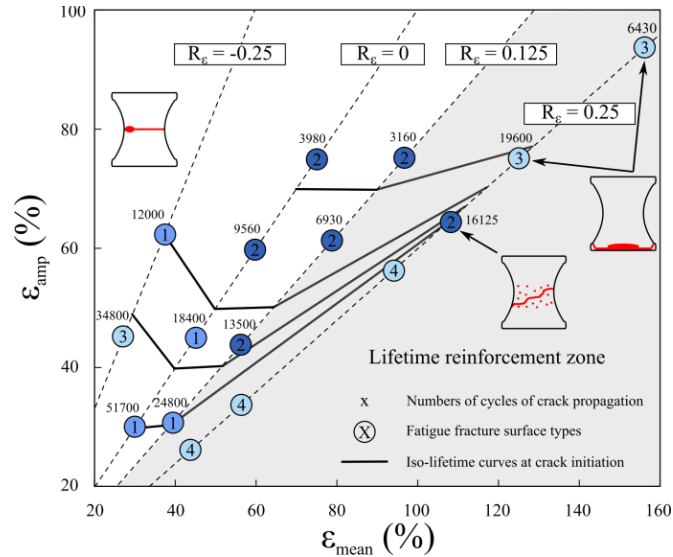

Figure 3: effect of the loading conditions on the fatigue life, the number of cycle of crack propagation and the macroscopic damage.

*Link between fatigue striation and loading conditions*

Fatigue striations have been linked to the loading condition in order to bring new information on the role of SIC in the crack growth resistance of NR. Classically, lifetime reinforcement occurs for non-relaxing loadings (*i.e.* R > 0), which can be represented in a Haigh diagram [12,13]. **Erreur ! Source du renvoi introuvable.** presents such a diagram obtained after fatigue tests at 23°C. The iso-lifetime curves in black highlight the lifetime reinforcement since their slope become positive under non-relaxing loadings (see [3] for more details). At the macroscopic scale, three fatigue damage modes are identified and reported in **Erreur ! Source du renvoi introuvable.**:
- crack initiation in the surface vicinity and crack propagation in the median section until total failure,
- crack initiation in the surface vicinity and propagation of the crack by bifurcation in the median section zone,
- crack initiation and growth close to the insert area.

The fact that the two later damage modes are obtained under non-relaxing loading highlights the strong influence of the loading on the damage mechanisms.

At the microscopic scale, three fracture surface types are identified. They are described as follows:
- ① fracture surface with Regime 1 fatigue striation,
- ② fracture surface with Regime 2 fatigue striation. Since Regime 2 is consecutive to Regime 1, fracture surfaces noted ② also exhibit Regime 1 fatigue striations,
- ③ fracture surface without any fatigue striation.

It is to note that fracture surfaces denoted ④ correspond to experiments where no failure was observed after 1.5 million cycles. The results are described in terms of fracture surface type, according to the loading ratios $R_\varepsilon$:

- for repeated tension ($R_\varepsilon = 0$), fatigue striations are observed independently of the maximal strain applied: for low $\varepsilon_{max}$, Regime 1 appears, Regime 2 occurs as $\varepsilon_{max}$ increases. Both striation height and distance between two striations increase with $\varepsilon_{max}$. It is in good agreement with the study by Munoz *et al.* [10].
- for non-relaxing tension ($R_\varepsilon > 0$), the same comment can be drawn as for repeated tension, however, Regime 2 appears for lower values of $\varepsilon_{max}$. Such loadings were found to promote SIC [3,14] since the loading never goes back to zero during a fatigue cycle. This could therefore explain the more important occurrence of Regime 2. Note that the fact that no striation is observed for the highest $\varepsilon_{max}$ is explained by the difference in terms of damage mechanisms: indeed, the initiation that occurred under the insert area did not allow the formation of striation.
- for relaxing tension ($R_\varepsilon > 0$), fatigue striations are also observed. However, they form at higher $\varepsilon_{max}$. It can be due to the fact that the crystallinity goes back to zero during such fatigue tests.

Finally, literature suggests that no crystallites remain at 90°C under static loadings [15]. Fatigue tests performed at 90°C revealed that no wrenchings nor fatigue striation form under such temperature. Therefore, it confirms the strong role of SIC on the formation of wrenchings and fatigue striation.

*Discussion*

The results presented in the present paper suggest that fatigue striations form whatever the loading applied is, even for tension-compressions tests where no lifetime reinforcement is observed. Therefore, it can be concluded that fatigue striations are the signature of SIC, but not the signature of the lifetime reinforcement. The fact that no striation was observed after fatigue tests carried out at 90°C could indicate that striation only form when a certain crystallinity threshold is reached. It has been shown that fatigue striation form under a wide range of loading: only Regime 1 occurs for relaxing loadings, for superior loading ratios (*i.e.* $R \geq 0$) the striation formation is found to be amplified as the loading ratio is increased. This result can be analyzed with respect to crack growth rate curves measured at different loading ratios in the study proposed by Lindley [1]. The authors showed that: (*i*) increasing the tearing energy (*i.e.* the loading) induces an increase of the crack growth rate: (*ii*) increasing the loading ratio decreases the crack growth rate, for a given tearing energy and increases the tearing energy threshold below which no crack growth happens. The author deduced from this result that the crack growth resistance of NR was due to SIC. Our results are in good agreement with Lindley's for non-relaxing tension ($R = 0.25$): below a certain maximum of loading, no crack propagation occurs (fracture surface ④), above this maximum of loading, only Regime 2 striation form. Based on the result presented by Lindley, the crack growth rate at which Regime 2 occurs for $R = 0.25$ loadings should be lower than the one at which it occurs for lower loading ratios. To verify this result, the number of cycles of crack propagation $N_p$ is also reported in **Erreur ! Source du renvoi introuvable.**. Since the tests are carried out at approximately the same strain rate, it can be considered that $N_p$ evolves with the inverse of the crack growth rate. Therefore, for a given amplitude of loading, since $N_p$ increases with the loading ratio in the lifetime reinforcement zone, the crack growth rate decreases. Our results suggest that under non-relaxing loadings, the loading ratio governs the striation typology. Furthermore, the crack propagation curves can be a useful tool to quantitatively analyse the signature of SIC on the failure surface.

## CONCLUSION

The role of SIC in the fatigue crack growth of NR is investigated using post-mortem analysis of fatigue striations. Two striation regimes occur: Regime 1 corresponding to striation patches with different orientations and Regime 2 corresponding to well defined striations. Striations are observed whatever the loading ratio applied and even for relaxing loadings, where no lifetime reinforcement occurs. Fatigue striations are therefore not the signature of the lifetime reinforcement. Nevertheless, the fatigue striation formation in Regime 2 seems to be encouraged for non-relaxing loadings, where the lifetime reinforcement occurs. For $R \leq 0$ loadings, increasing the loading (*i.e.* the tearing energy) increases the crack growth rate, which is in good agreement with the literature. For $R > 0$, the striation typology evolution could be governed by the loading ratio. Finally, no striation was observed when the temperature was increased to 90°C. Therefore, it is concluded that the formation of striation requires the crystallinity to be above a crystallinity threshold under which no striation form.

## ACKNOWLEDGEMENTS

The authors thank the Cooper Standard France company for supporting this work and for fruitful discussions. The authors thank also the National Center for Scientific Research (MRCT-CNRS and MI-CNRS) and Rennes Metropole for supporting this work financially. SEM images were performed at CMEBA facility (ScanMAT, University of Rennes 1), which received a financial support from the European Union (CPER-FEDER 2007-2014).